\documentstyle[preprint,prl,aps,epsfig]{revtex}
\begin{document}
\preprint{LBL-PUB-45566, May, 2000} 

\title{Localized Beampipe Heating due to $e^-$ Capture and Nuclear
Excitation in Heavy Ion Colliders}

\author{Spencer R. Klein} 
\address{Lawrence Berkeley National Laboratory, Berkeley, CA 94720} 

\break 
\maketitle
\vskip -.2 in
\begin{abstract}
\vskip -.2 in 

At heavy ion colliders, two major sources of beam loss are expected to
be $e^+e^-$ production, where the $e^-$ is bound to one of the nuclei,
and photonuclear excitation and decay via neutron emission.  Both
processes alter the ions charged to mass ratio by well defined
amounts, creating beams of particles with altered magnetic rigidity.
These beams will deposit their energy in a localized region of the
accelerator, causing localized heating, The size of the target region
depends on the collider optics.  For medium and heavy ions, at design
luminosity at the Large Hadron Collider, local heating may be more
than an order of magnitude higher than expected.  This could cause
magnet quenches if the local cooling is inadequate.  The
altered-rigidity beams will also produce localized radiation damage.
The beams could also be extracted and used for fixed target
experiments.


keywords: ion colliders; beam loss; local heating; beam extraction;
electron capture; giant dipole resonance.

\end{abstract}
\narrowtext

\section{Introduction}

Ion colliders are expected to lose beam particles via ion-ion
collisions.  Besides purely hadronic interactions, photonuclear
interactions are of considerable interest\cite{intro}\cite{baur} and
many electromagnetic processes have cross sections larger than the
hadronic cross section.  Two electromagnetic processes are expected to
be major sources of beam particle loss: production of $e^+e^-$ pairs
by the colliding electromagnetic fields, where the electron is
produced bound to one of the nuclei, and the excitation of one nucleus
by the electromagnetic field of the other.  In the latter case, the
nucleus will be excited to a Giant Dipole Resonance (GDR) or higher
state.  Usually, the GDR decays by emitting one or more neutrons.

Both of these interactions alter the mass to charge ratio (rigidity)
of the affected ion.  For the heavy ions like gold or lead, the
rigidity increases about 1\% for electron capture, and decreases about
0.5\% for neutron loss.  For lighter ions, the change is larger.
Because these changes are larger than the acceptance of the magnetic
optics, these ions are lost from the beam, and eventually strike the
beam pipe.  The hadronic showers from these collisions will deposit
their energy in the cryogenic magnets around the beampipe.  Averaged
over the entire ring, the energy deposition is small.  However,
because of the well defined rigidities, the target area is a small
fraction of the ring, and localized heating and radiation damage may
be a problem.  The altered rigidity beams could also be extracted from
the collider and used as a test beam.
 
Here, we consider these reactions for the Relativistic Heavy Ion
Collider (RHIC) at Brookhaven National Laboratory\cite{RHICCDR} and
the Large Hadron Collider (LHC) now under construction at
CERN\cite{ALICE}.  Table~\ref{lumin} lists energies and luminosities
for the systems considered here.  For the LHC, different sources quote
somewhat different luminosities.  This paper will use the peak
(initial) luminosities given in the RHIC conceptual design report and
the ALICE proposal for the LHC.  The LHC luminosities are for 125 nsec
bunch spacing and collisions in 1 experimental hall\cite{ALICE}.  If
the bunch spacing is decreased to 25 nsec, the luminosity increases by
a factor of 5.

\section{$e^+e^-$ production and $e^-$ capture}

The electromagnetic fields of the colliding nuclei may interact and
produce $e^+e^-$ pairs.  The electron can be produced bound to one of
the nuclei, reducing the net charge by $1$; usually the electron is
captured by the $K$-shell.

The cross section for pair production and electron capture can be
calculated using a number of techniques.  Although some coupled
channel calculations have found very high cross sections, new
all-orders analytic calculations support the results of perturbative
calculations\cite{baltzem}, despite the large coupling constant,
$Z\alpha\sim 0.6$.  For beams of identical nuclei, the cross section
for capture to a $K$-shell by a charge $Z$ nucleus is\cite{firstrev}
\cite{baur}
\begin{equation}
\sigma(A + A\rightarrow Ae^- + A + e^+) = 
{33\pi Z^8\alpha^6r_e^2 \over 10}
{1\over e^{2\pi\alpha Z} -1 } 
\bigg[\ln{({\delta(\gamma^2-1)\over2})}-{5\over 3}\bigg]
\end{equation}
where $\alpha=e^2/\hbar c$ is the fine structure constant, $r_e$ the
classical electron radius, $\gamma$ the Lorentz boost of a single
beam, and $\delta\sim 0.681$.  This is the cross section to excite a
specific nucleus; the cross section to excite either nucleus is twice
as large.

The effect of inclusion of higher shells is to boost this cross
section by 20\%\cite{firstrev}. With this correction, the cross
sections are given in Table~\ref{lumin}; the cross sections drop
dramatically as $Z$ decreases; the energy dependence is moderate.

A recent extrapolation from lower energy data\cite{grafstrom} found
higher cross sections, 94 barns for gold at RHIC and 204 barns for
lead at the LHC, twice the perturbative result. If this result is
correct, then heating will be twice that predicted.

The particle loss rates, the products of these cross sections and the
luminosity, are given in Table~\ref{losses}.  The very strong $Z$
dependence of the cross section is compensated by the rapid luminosity
increase as $Z$ decreases, and the particle losses are largest for
medium ions.  Table~\ref{losses} also gives the single beam energy
losses, which scales with the atomic number $A$.  Because the escaping
positron has a very small momentum, the tiny nuclear momentum change
will not affect the rigidity.

\section{Nuclear Excitation}

Many types of electromagnetic excitation are possible for nuclei.
Single or multiple photon absorption is possible, and the excited
states can decay via single or multiple neutron emission or by nuclear
breakup.  The most common is a Giant Dipole Resonance (GDR), where the
protons and neutrons oscillate collectively against each other.  GDRs
usually decay by single neutron emission.  Because of its large cross
section, we focus on GDR excitation and single neutron decay.  Higher
resonances typically have more complex decays, often involving
multiple neutrons.

The cross section for photonuclear excitation of a given nucleus is
\begin{equation}
\sigma(A + A \rightarrow A_{GDR} + A) = 
\int_0^\infty {dn_\gamma\over dk} \sigma_{GDR} (k) dk
\label{egda}
\end{equation}
where $k$ is the photon energy, $dn_\gamma/dk$ is the
Weizs\"acker-Williams photon spectrum, subject to the condition that
the nuclei don't interact hadronically\cite{baur}\cite{ussra}, and
$\sigma_{GDR}(k)$ is the GDR excitation cross section. For heavy
nuclei, the cross section peaks $k=31.2 A^{-1/3} {\rm\ MeV} + 20.6
A^{-1/6}$ MeV\cite{rmp}. For symmetric systems, the cross section can
be parameterized\cite{gdr}
\begin{equation}
\sigma(A + A \rightarrow A_{GDR} + A) = 
3.42 \mu b {(A-Z)Z^3\over A^{2/3} } \ln{(2\gamma^2-1)}.
\label{egdr}
\end{equation}

Because this formula does not include other photoexcitation processes
this cross section is lower than some quoted elsewhere. For example,
Ref.~\cite{ALICE} gives a formula for GDR electromagnetic dissociation
with a coefficient about 20\% higher than Eq.~(\ref{egdr}).  The cross
sections from Eq.~(\ref{egdr}), listed in Table~\ref{lumin}, are less
well determined than for electron capture, because of competition with
higher order processes, which can lead to multiple neutron emission.
For lower energy gold interactions, the cross section for 2-neutron
emission is about 20\% of single neutron emission cross
section\cite{hill}.  Of course, this ratio could rise at higher
energies.  For lighter nuclei, the higher order processes should be
much less important.  Since Eq.~(\ref{egdr}) fits existing low energy
data fairly well, we use it here.

For heavy ions, these cross sections are comparable to those for
electron capture.  Table~\ref{losses} shows single beam loss rates for
GDR excitation.  For heavy ions, losses are comparable with electron
capture. Because GDR excitation is much less $Z$-dependent than
electron capture, it is the dominant process for lighter ions.  Losses
range up to 1.6 million particles/second and 36 watts for calcium at
the LHC.

Unlike electron capture, with neutron emission, the nuclear recoil is
significant.  The $\gamma + A\rightarrow (A-1) +n$ reaction is a
two-body problem; neglecting the small change in binding energy, in
the rest frame the nuclear recoil momentum is $p_{exc.}=\sqrt{2m_n
k}$, where $m_n$ is the nucleon mass.  In the lab frame, the nuclear
momentum change is a maximum of $\Delta p/p=p_{exc.}/Am_n$, depending
on the emission direction.

\section{Target Region}

Because the rigidity change is small, the affected nuclei will strike
the beampipe a considerable distance down the beampipe, with the exact
location depending on the beam optics.  Here, we consider a simple
model, with the nuclei in circular orbits in a constant magnetic
field.  The radius of curvature is $R=p_A/ZecB$, where $p_A$ is 
nuclear momentum and $B$ the magnetic field. For an intact
nucleus, $R$ must match the accelerator radius $R_0$.

Electron capture decreases $Z$ by $1$, so the radius of curvature
increases to $R=Z/(Z-1)R_0$, a change of $\Delta R = R_0/Z$.  The
trajectory will gradually be displaced outward from the beampipe
centerline, with the displacement from the center growing as the
square of the distance travelled.  If the nucleus travels an angle
$\theta$ around the accelerator, the displacement is $x=2\Delta
R\theta^2/\pi^2$.  It will strike a beampipe with radius $d$ after
moving an angle $\sqrt{\pi d/2\Delta R}$, a distance
\begin{equation}
D_c = \theta R_0 = \pi \sqrt{R_0Zd\over 2}
\end{equation}
downstream from the interaction region.  In the RHIC magnets, $d=3.45$
cm \cite{RHICCDR} while for LHC the horizontal magnet opening is
$d=2.2$ cm\cite{LHCvacuum}. Values of $D_c$ are listed in
Table~\ref{target}.

Neutron loss decreases the rigidity by a factor $(A-1)/A$, reducing
the radius of curvature to $R=(A-1)/A R_0$.  With $\Delta R=R_0/A$,
the nucleus will hit the beampipe a distance
\begin{equation}
D_d = \pi \sqrt{R_0Ad\over 2}
\end{equation}
downstream. Table~\ref{target} gives values of $D_d$.  Since $A\sim
2Z$, $D_d \sim \sqrt{2} D_c$.  

$D_c$ and $D_d$ are the distances to the middle of the target regions;
the length of the target region depends on the spreading of the
altered beam.  Several factors contribute to the spreading: the
momentum spread of the incident beam, nuclear recoil (for GDR
excitation), and the size of the hadronic shower when the particle
hits the beampipe.  Focusing from the beam optics will also have a big
effect; this factor is neglected here.  We simply assume that the
optics remove the effect of perpendicular momentum variations, leaving
the longitudinal momentum variations unaffected.  We will add the rms
momentum spreading for each of these factors in quadrature, neglecting
corrections due to the non-Gaussian nature of the distributions.

RHIC is designed for a maximum momentum spread, $\Delta p/p=1.5\times
10^{-3}$\cite{RHICCDR}\cite{peggs}.  At the LHC $\Delta
p/p=10^{-4}$\cite{pdg}.  These $\Delta p/p$ are maximum variation;
$\sigma p/p \approx (1/\sqrt{3})\ \Delta p/p$.  These numbers are
typical; with time, intra-beam scattering and beam-beam interactions
increase the momentum spread.

The momentum spread also affects the radius of curvature, with $\sigma
R/R_0 = \sigma p/p$.  Neglecting magnetic focusing, an intact beam
particle with individual momentum variation $\delta$ will hit the
beampipe a distance
\begin{equation}
D_\delta = \pi \sqrt{ Rd \over 2 (|\delta|/p)}
\end{equation}
downstream from the interaction region.  Without magnetic focusing,
particles with $\delta/p = \sigma p/p$ would strike the beampipe 118
meters and 8.5 km downstream at RHIC and LHC respectively.

For GDR excitation, the recoil from the neutron emission affects the
ion momentum.  $p_{exc.}$ is the maximum momentum change; this can be
approximated as a Gaussian with $\sigma p/p \approx 0.6 p_{exc}/p_A$.
This $\sigma p/p$ is added in quadrature to the beam spread.  It is
usually the dominant factor.

The interactions and momentum changes are combined by adding the
$\sigma R$, with
\begin{equation}
D_\pm = \pi R \sqrt{d\over 2\sigma R_\pm}
\end{equation}
where $\sigma R_\pm$ are found by adding and subtracting the $\sigma
R$ due to momentum spread from the $\sigma R$ from the rigidity
change.  Energy is deposited over a length $L=D_+ - D_-$.  Assuming
that the individual particle momenta follow a Gaussian distribution,
68\% of the particles hit within this target area.  These $L$, given
in Table~\ref{target}, are small compared to $D_\pm$.  The small $L:D$
ratio is an indication that magnetic focusing will not drastically
change the picture presented here.

When the nucleus hits the beampipe, the size and shape of the hadronic
shower depend on the target geometry and magnetic field. This will
affect how much of the energy is deposited in the cryogenically cooled
magnet.  The magnet assembly can be approximated as
copper\cite{quencha}, which has a hadronic interaction length
$\Lambda=15$ cm.  At 100 GeV/nucleon (3.5 TeV/nucleon), 99\% (95\%) of
the energy is deposited within $10\Lambda$\cite{pdg}, or 1.5 meters.
Most of this energy is deposited within a few $\Lambda$ of the point
of maximum shower development; we treat the energy deposition as a
Gaussian, with $2\sigma=6\Lambda$=0.9 meters.  This $2\sigma$ is added
in quadrature with $L$ to give the total target length.  So, 68\% of
the total energy should be deposited within this region.

Of course, some of the energy will escape down the beampipe or into
the surrounding environment.  Here, we assume that half of the energy
reaches the cold volume, with the other half escaping.  With these
assumptions, the average power dissipations are given in
Table~\ref{results}.  Even though the electron capture and GDR beams
have similar power, the energy deposition is more localized for
electron capture because of the GDR nuclear recoil. Lead and niobium
are the most problematic, depositing 2.6 W/m and 3.2 W/m respectively.

These loads must be compared with the local cooling capacity.  The
RHIC Conceptual Design Report does not specify a value for beam
induced heating loads.  However, at 4$^o$K, 2.5 Watts of cooling is
planned for a 9.7 meter long dipole\cite{RHICCDR}.  The power
dissipations are far smaller than this.

At the LHC, the expected heat loads are much higher.  The main
sources, synchrotron radiation and image currents are expected to
deposit 0.6 W/m and 0.8 W/m on the beampipe respectively.  A screen
will be installed inside the magnets to divert this heat from the
$1.9^o$K magnets\cite{LHCchallenges}; less than 0.1 W/m is expected to
leak through the screen.  The accelerator design also allows for 0.1
W/m from inelastic nuclear scattering which cannot be
shielded\cite{LHCvacuum}, for a total of 0.2 W/m.  Because of the low
luminosity, synchrotron radiation and image currents are negligible
for ion collisions, so the entire 0.2 W/m could be 'allocated' to beam
losses.  This 0.2 W/m is less than 10\% of the 2.1 and 3.0 W/m
expected from electron capture for lead and niobium beams.

The local temperature rise from this energy will depend on the local
cooling capacity and thermal resistance.  At 7 TeV, a loss of $8\times
10^6$ protons/meter/second will induce a quench\cite{quencha}; this is
about 8 watts/meter, uncomfortably close to the heat loads calculated
above.

Since the altered-rigidity beams will remain in the horizontal plane
of the accelerator, and strike the outside (for electron capture) and
inside (for GDR excitation) of the beampipe, the heating will be
uneven, and local hot spots are likely.  These hot spots could induce
a quench even if the average power is below the quench limit.

\section{Discussion}

At RHIC, the local heating due to altered-rigidity beams is within the
available cooling capacity.  At the LHC, these altered rigidity beams
have higher powers, up to 36 watts.  At the same time, the target
regions are shorter than at RHIC, and the cooling capacities are
somewhat lower, because the LHC uses supercooled magnets.  With
niobium beams, the expected heating is 3.2 W/m, 16 times the expected
beam heat load of 0.2 W/m.  For lead, the 'standard' ion choice, the
heating is 2.6 W/m, 13 times the expected load.  

These loads are close to the expected quench limit of 8 W/m.  When the
detailed distribution of energy deposition is considered, electron
capture from either lead or niobium beams might deposit enough energy
to cause a magnet quench.  With GDR, the heat loads are lower, but may
be problematic for niobium and calcium beams.

These estimates are based on back-of-the-envelope calculations; the
uncertainties are correspondingly large.  The most important missing
factors are the charged particle optics and the magnet arrangement.
The former could change the pattern of the energy deposition, by
decreasing or increasing the length of the target area, while the
latter determines how the energy affects the magnet.  Detailed
simulations are needed to study both factors.  At the LHC, a
calculation using the magnetic dispersion finds target regions with
$L=1$ m for lead, very close to that found here\cite{bernard}.

However, for lead and niobium, it is not unlikely that supplementary
cooling will be required.  Alternately, it might be possible to
install new collimators to channel the energy deposition.  Current LHC
plans call for a single collimator, located in one of the interaction
regions\cite{LHCcollimator}.

If the larger estimates of electron capture cross section are correct,
the margin for error would almost completely disappear.  Alternately,
in higher-luminosity scenarios, the heating can be up to five times
higher.  In these scenarios, localized energy deposition could be the
luminosity limiting factor.

The beams will also deposit significant radiation in these regions.
Although radiation damage is beyond the scope of this article, current
studies neglect this source\cite{radiate}.

Finally, it might be possible to extract these altered-rigidity beams
and use them for fixed target experiments. Electron capture beams are
the most appropriate for extraction, because of the smaller emittance.
The particle rates are comparable to those at existing fixed target
heavy ion accelerators.  At the LHC, the beam energies would exceed
existing fixed target sources.

\section{Conclusions}

Both pair production with capture and nuclear excitation with neutron
emission produce beams of ions with altered magnetic rigidity.  These
beams will follow defined trajectories and strike the collider
beampipe downstream of the interaction regions, producing localized
energy deposition.

At RHIC, this energy deposition is small, and should not cause
problems. However, with lead, niobium or calcium beams at the LHC,
simple calculations indicate that the energy deposition will be far
larger than the planned cooling.  Further studies are needed to
confirm these simple models.  If the local heating exceeds the
available cooling, the magnets could quench; electron capture could
limit the luminosity achievable with heavy or medium ions. In
addition, these beams could cause localized radiation damage.  These
problems become even worse for high luminosity running, with 25 nsec
spacing between heavy ion bunches.

On the positive side, with appropriate optics, these beams could
be extracted and used for fixed target experiments.

It is a pleasure to acknowledge useful exchanges with Bernard
Jeanneret, Nikolai Mokhov and Steve Peggs.  This work was supported by
the US DOE, under contract DE-AC-03-76SF00098.

\begin{table}
\begin{tabular}{lrrrrr}
Machine & Ion & Beam Energy & Design Luminosity 
& $\sigma$ ($e^-$ capture) & $\sigma (GDR)$ \\
\hline
RHIC    & gold  	& 100 GeV/n     & $2\times10^{26}cm^{-2}s^{-1}$ &
45 b     & 58 b    \\
RHIC    & iodine  	& 104 GeV/n     & $2.7\times10^{27}cm^{-2}s^{-1}$ &
6.5 b     & 15 b    \\
RHIC	& silicon  	& 125 GeV/n     & $4.4\times10^{28}cm^{-2}s^{-1}$ &
1.8 mb    & 150 mb  \\
LHC     & lead   	& 2.76 TeV/n    & $1\times10^{27}cm^{-2}s^{-1}$ &
102 b     & 113 b   \\
LHC     & niobium   	& 3.1 TeV/n     & $6.5\times10^{28}cm^{-2}s^{-1}$ &
3.1 b     & 10 b    \\
LHC     & calcium   	& 3.5 TeV/n     & $2\times10^{30}cm^{-2}s^{-1}$ &
36 mb    & 800 mb  \\
LHC     & oxygen   	& 3.5 TeV/n     & $3\times10^{31}cm^{-2}s^{-1}$ &
81 $\mu$b & 37 mb   \\
\end{tabular}
\caption[]{Luminosity and beam kinetic energy for heavy ion beams at
RHIC and LHC.  The RHIC luminosities are from Ref.~\cite{RHICCDR}.
Different references quote somewhat different ion luminosities for the
LHC; these are the peak luminosities for a single experiment and 125
nsec bunch spacing from Table 8.3 of Ref. \cite{ALICE}.  Also given
are cross sections for electron capture and for electromagnetic GDR
excitation followed by single neutron emission.}
\label{lumin}
\end{table}

\begin{table}
\begin{tabular}{lrcrr}
Beam  & Capture Loss (pps) & Capture Power Loss  & 
GDR loss (pps) &  GDR Power loss  \\
\hline
RHIC - Au &    8,900 & 28 mW    &    12,000   & 37 mW \\
RHIC - I &    18,000 & 37 mW    &    40,000   & 85 mW \\
RHIC - Si &       82 & 46$\mu$W &     6,500   & 3.6 mW\\
LHC- Pb &    102,000 & 10. W    &   113,000   & 11 W \\
LHC- Nb &    196,000 & 9.1 W    &   650,000   & 30 W  \\
LHC- Ca &     73,000 & 1.6 W    & 1,600,000   & 36 W  \\
LHC- O &       2,400 & 22 mW    & 1,100,000   & 10 W   \\
\end{tabular}
\caption[]{Single Beam loss rates, in particles per second (pps), and
single beam power losses, for electron capture and GDR excitation.}
\label{losses}
\end{table}

\begin{table}
\begin{tabular}{lrrr}
Accelerator  & Mean Accelerator Radius  & Beam Pipe Radius 
& $\Delta p/p$ \\ 
\hline
RHIC &  610 m   & 3.45 cm  & $1.5\times 10^{-3}$ \\
LHC   & 4245 m  & 2.2 cm  & $1\times 10^{-4}$ \\
\end{tabular}
\caption[]{Some characteristics of RHIC and LHC.  The accelerators are
not perfectly circular; the radius is calculated from the overall
circumference.  The beampipe radius is the horizontal aperture inside
the magnets.  The momentum spread is somewhat ion-dependent.  The
spread will increase gradually as the beams circulate and collide.}
\label{accel}
\end{table}

\begin{table}
\begin{tabular}{lrrrr}
Beam  &  Capture-Distance  & Capture $L$ 
& GDR Distance & GDR $L$ 
\\
\hline
RHIC - Au & 91 m      & 6.2 m     &  143 m     & 28 m \\
RHIC - I  & 74 m      & 3.4 m     &  115 m     & 18 m  \\
RHIC - Si & 38 m      & 0.5 m     &   54 m     & 7 m\\
LHC- Pb   & 194 m     & 0.9 m     &  309 m     & 31  m \\
LHC- Nb   & 137 m     & 0.3 m     &  207 m     & 22 m \\
LHC- Ca   & 96 m      & 0.11 m    &  136 m     & 16  m \\
LHC- O    & 60 m      & 0.03 m    &   86 m     & 12 m\\
\end{tabular}
\caption[]{Beam impact point (distance from the interaction region)
for electron capture and GDR excited nuclei, based on a simple
geometric model.  The $L$ columns show the variation in impact
distance due to the beam momentum spread and, for GDR excitation,
including nuclear recoil; the recoil usually dominates over the
accelerator energy spread.  The hadronic shower development is not
included.}
\label{target}
\end{table}

\begin{table}
\begin{tabular}{lrrrr}
Beam & Capture Power Dissipation &
GDR Power Dissipation & Cooling Capacity \\
\hline
RHIC - Au	& 1.5 mW/m 	& 0.4 mW/m & -  \\
RHIC - I 	& 3.6 mW/m 	& 1.6 mW/m &  - \\
RHIC - Si	& 15 $\mu$W/m 	& 0.2 mW/m &  - \\
LHC - Pb	& 2.65 W/m  	& 0.1 W/m & 0.2 W/m \\
LHC - Nb	& 3.2 W/m  	& 0.5 W/m & 0.2 W/m \\
LHC - Ca	& 0.6 W/m  	& 0.7 W/m & 0.2 W/m \\
LHC - O		& 82 mW/m 	& 0.3 W/m & 0.2 W/m \\
\end{tabular}
\caption[]{Power dissipation per unit length for electron capture and
GDR excitation.  Both dissipations assume 68\% of the power is
deposited within a $\pm 1\sigma$ target region, where the $\sigma$
includes the momentum spread, GDR recoil, and hadronic shower.  Half
of this energy is assumed to end up in the magnet, with the other half
escaping.  Also shown, for comparison, is the planned cooling capacity
at the LHC.  If the LHC uses a 25 nsec bunch spacing to increase the
luminosity, the dissipation will grow by a factor of 5.}
\label{results}
\end{table}

\begin{references}
\def\etal{{\it et al.}}

\bibitem{intro} J. Nystrand and S. Klein, nucl-ex/9811007, in {\it
Proc. Workshop on Photon Interactions and the Photon Structure}, eds.
G. Jarlskog and T. Sj\"ostrand, Lund, Sweden, Sept., 1998.

\bibitem{baur}G. Baur, K. Hencken and D. Trautman, J. Phys. {\bf G24}
(1998) 1457.

\bibitem{RHICCDR}{\it Conceptual Design of the Relativistic Heavy Ion
Collider}, BNL-52195, May, 1989, Brookhaven National Laboratory.

\bibitem{ALICE}N. Ahmad \etal, {\it ALICE Technical Proposal},
CERN/LHCC 95-71, Dec., 1995.  These calculations and results are also
in D. Brandt, K. Eggert and A. Morsch, CERN AT/94-05, March, 1994.

\bibitem{baltzem}A. J. Baltz, M. J. Rhoades-Brown and J. Weneser,
Phys. Rev. A {\bf 47} (1993) 3444.

\bibitem{firstrev}C. A. Bertulani and G. Baur, Phys. Rep. {\bf 163}
(1988) 299.

\bibitem{grafstrom}P. Grafstr\"om \etal, {\it Measurement of
Electromagnetic Cross Sections in Heavy Ion Interactions and its
consequences for Luminosity Lifetimes in Ion Colliders},
CERN-SL-99-033 EA.

\bibitem{ussra}S. Klein and J. Nystrand, Phys. Rev. {\bf C60},
014903 (1999).

\bibitem{rmp}B. L. Berman and S. C. Fultz, Rev. Mod. Phys.
{\bf 47} (1975) 713.

\bibitem{gdr}G. Baur and C. A. Bertulani, Nucl. Phys. {\bf A505}
(1989) 835.

\bibitem{hill}T. Aumann \etal, Phys. Rev. {\bf C47} (1993) 1728.

\bibitem{LHCvacuum}O. Gr\"obner, LHC Project Report 181, May, 1998,
presented at the 1997 Particle Accelerator Conference, Vancouver,
Canada, May, 1997.

\bibitem{peggs} S. Peggs (private communication, 2000) quotes an
$\sigma p/p = 8\times 10^{-4}$ at the end of a spill; this is
slightly lower than the $\Delta p/p$ used here.

\bibitem{pdg}C. Caso \etal (PDG group), Eur. Phys. J. {\bf C3} (1998)
1.

\bibitem{quencha}J. B. Jeanneret, D. Leroy, L. Oberli and T.
Trenkler, {\it Quench Levels and Transient Beam Losses in LHC
Magnets}, LHC Project Report 44, July, 1996.

\bibitem{LHCchallenges}L. R. Evans, {\it LHC Accelerator Physics and
Technology Challenges}, LHC Project Report 303, April, 1999, presented
at the 1999 Particle Accelerator Conference, March 29-April 2, 1999,
New York, USA.

\bibitem{radiate}S. Roesler and G. R. Stevenson, {\it Estimation of
Dose to Components close to the low-$\beta$ insertions at LHC point 2
(ALICE),} CERN/TIS-RP/IR/99-15, April, 1999.

\bibitem{bernard}B. Jeanneret, private communication, 2000.

\bibitem{LHCcollimator}A. I. Drozhdin and N. V. Mokhov, {\it
Optimisation of the LHC Beam Cleaning System with Respect to Beam
Losses in the High Luminosity Insertions}, CERN-LHC-Project-Report
148, Oct., 1997; I.Azhgirey, I. Baishev, N. Catalan Lasheras, J. B.
Jeanneret, {\it Cascade Simulations for the LHC Betatron Cleaning
Insertion}, CERN-LHC-Project-Report 184, May, 1998.

\end{references}
\end{document}